\documentclass[conference]{IEEEtran}
\IEEEoverridecommandlockouts
\usepackage{hyperref}
\usepackage{url}
\usepackage{subfig}
\usepackage{cite}
\usepackage{amsmath,amssymb,amsfonts}
\usepackage{algorithmic}
\usepackage{graphicx}
\usepackage{textcomp}
\usepackage{xcolor}
\def\BibTeX{{\rm B\kern-.05em{\sc i\kern-.025em b}\kern-.08em
    T\kern-.1667em\lower.7ex\hbox{E}\kern-.125emX}}
\begin{document}

\title{Extracting and Visualizing Wildlife Trafficking Events from Wildlife Trafficking Reports}

\author{Devin Coughlin\texttt{$^\bigstar$},
	Maylee Gagnon\texttt{$^\bigstar$},
	Victoria Grasso\texttt{$^\bigstar$},
	Guanyi Mou\texttt{$^\bigstar$},
	Kyumin Lee\texttt{$^\bigstar$},
	Renata Konrad\texttt{$^\bigstar$},\\
    Patricia Raxter\texttt{$^\diamondsuit$},
	Meredith Gore\texttt{$^\heartsuit$}\\
	\texttt{$^\bigstar$}Worcester Polytechnic Institute,
	\texttt{$^\diamondsuit$}Focused Conservation,
	\texttt{$^\heartsuit$}University of Maryland \\
	\texttt{\{djcoughlin,mrgagnon,vygrasso,gmou,kmlee,rkonrad\}@wpi.edu} \\
	\texttt{praxter@focusedconservation.org} \\
	\texttt{gorem@umd.edu}  \\
}


\maketitle

\begin{abstract}
Experts combating wildlife trafficking manually sift through articles about seizures and arrests, which is time consuming and make identifying trends difficult. We apply natural language processing techniques to automatically extract data from reports published by the Eco Activists for Governance and Law Enforcement (EAGLE). We expanded Python spaCy’s pre-trained pipeline and added a custom named entity ruler, which identified 15 fully correct and 36 partially correct events in 15 reports against an existing baseline, which did not identify any fully correct events. The extracted wildlife trafficking events were inserted to a database. Then, we created visualizations to display trends over time and across regions to support domain experts. These are accessible on our website, \href{https://wildlifemqp.github.io/Visualizations/}{Wildlife Trafficking in Africa}. 
\end{abstract}

\begin{IEEEkeywords}
wildlife trafficking, extraction, database, visualization
\end{IEEEkeywords}

\section{Introduction}
Wildlife trafficking, or the illegal wildlife trade, is a global issue that endangers populations of flora and fauna, ecosystems, and human security \cite{unodc2019}; the activity is known to occur in over 150 countries and territories and involves nearly 6,000 species \cite{unodc2020}. Wildlife trafficking poses widespread risks for species conservation, increases the spread of zoonotic disease \cite{emergingViralDiseases,smith2012zoonotic}, and can converge with other types of illicit crime such as drug and antiquities trafficking \cite{anagnostou2021illegal,uhm2020}. 
Most recently, social media platforms and e-commerce sites have facilitated a veritable explosion in wildlife trafficking, enabling wildlife traffickers to hide in plain sight, reaching consumers across the globe. Even where laws provide strong protections against wildlife trafficking, law enforcement, prosecutors, and judiciaries often fail to take wildlife trafficking seriously or lack resources to do so. Policing wildlife crime online rarely occurs for a variety of reasons. Linking evidence to action remains time consuming and tedious, so much so that authorities often remain many steps behind the high-level offenders they aim to arrest and deter; the sheer volume of disparate, unstructured, and disaggregated data makes it very difficult for scientists supporting decision-makers to identify trends and draw inferences.

\begin{figure}[ht]
\centering
    \includegraphics[width=\linewidth]{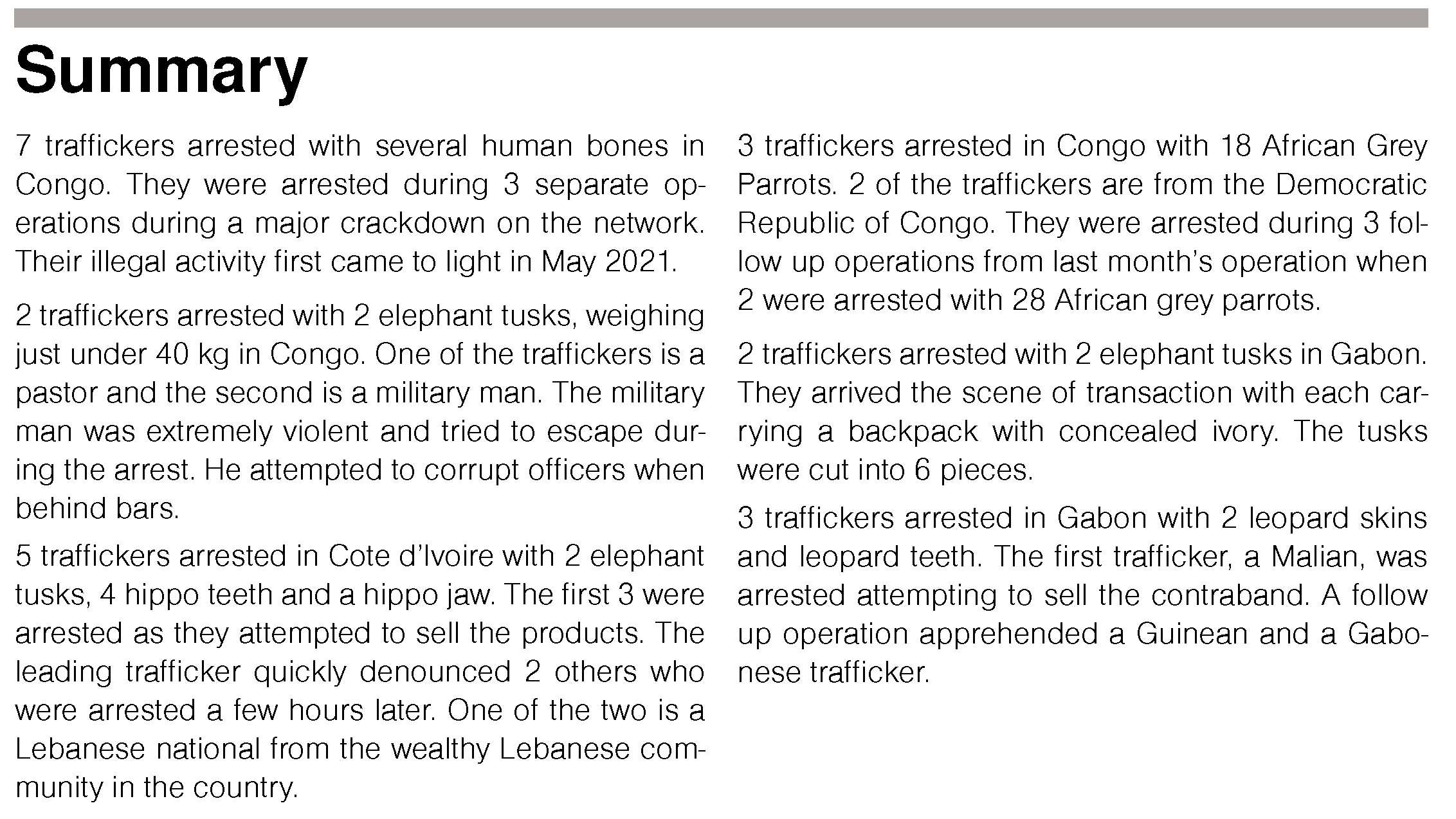}
    \caption{Excerpt of EAGLE Network Monthly Briefing in December 2021.}
    \label{fig:summary}
\end{figure}

Importantly, many nongovernmental organizations working to combat wildlife trafficking have open access websites, blog posts, newsletters, and briefings. The Eco Activists for Governance and Law Enforcement Network (EAGLE) \cite{eagle} is one such example. EAGLE aims to disrupt wildlife trafficking networks and related corruption through civic activism and collaborating with governments. The organization achieves its objectives through investigations, arrests, prosecutions, and publicity. The organization covers nine countries, Cameroon, Congo, Gabon, Togo, Senegal, Benin, Côte D’Ivoire, Burkina Faso, and Uganda. EAGLE posts monthly briefings detailing their activities (e.g., seizures, arrests). Such briefings have data about different species, countries, obfuscation methods, quantities and volumes, and modes of transportation (Figure~\ref{fig:summary}).

As a result of EAGLE’s activities, the organization estimates more than 2,000 wildlife traffickers have been jailed across West and Central Africa. If such data was aggregated and structured, it could (1) provide novel and high value information for investigators to identify trends and draw inferences about wildlife trafficking and (2) dramatically reduce the time investigators must invest in chasing data dead ends. To these ends, we designed and developed a novel natural language processing (NLP) pipeline to extract data from EAGLE reports using machine learning techniques and effectively display the data.

\begin{figure*}[ht]
\centering
    \includegraphics[width=\linewidth]{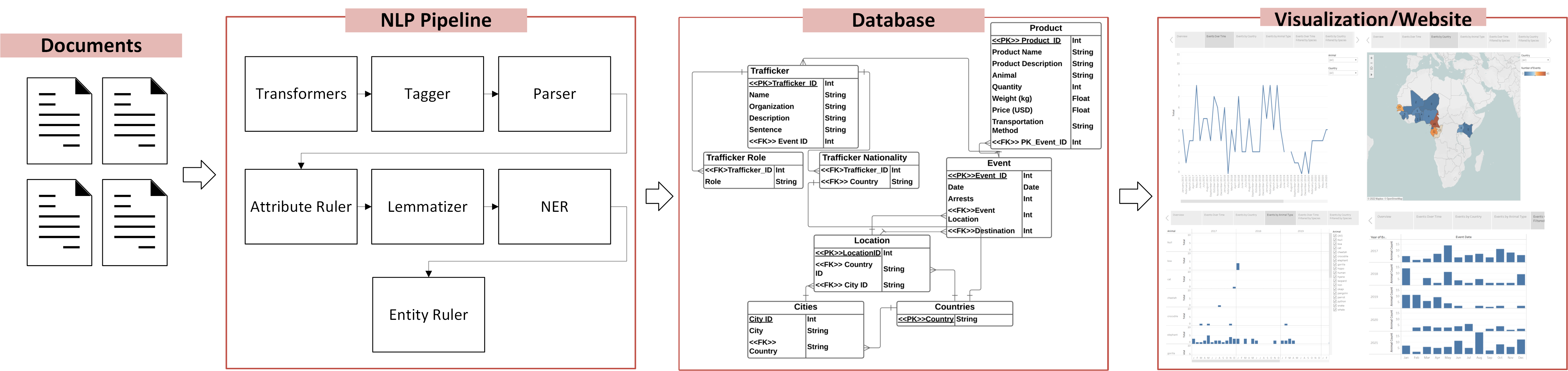}
    \caption{Our three-phased framework.}
    \label{fig:framework}
\end{figure*}

We present a framework built to extract wildlife trafficking-related named entities, unify data in a database, and visualize them in support of investigators seeking to understand trends of the events (e.g., species targeted, geo-graphic focus, and time of year). Our NLP pipeline for extracting wildlife trafficking events based on extracted named entities shows effectiveness of our approach, identifying 51 fully or partially correct events from 15 EAGLE reports.


\section{Related Work}
An early attempt to utilize NLP and machine learning algorithms to identify illegal wildlife trade activities captured in open-source reporting online: however, the process required a manual review of every report by an analyst \cite{sonricker2012digital}.  Miller et al. \cite{miller2019detecting} used machine learning along with web scraping and data visualization to create a six staged process to detect online transactions and sales activities involving any plant or animal listed on a Convention of Endangered Species of Wild Fauna and Flora (CITES) Appendix; however, it relied on using a specified species rather than the creation of a well-categorized database. NLP and machine learning have been used on Twitter data to analyze and classify posts with wildlife trafficking keywords \cite{xu2019use} to find code words and images related to the rhino horn trade \cite{nalluri2021survey}. Web scraping, NLP, deep learning, and visualization techniques can be used to collect information on illicit trade such as fake, counterfeit, and unapproved COVID-19-related health care products \cite{mackey2020big}. In our effort, we tied a specific machine learning technique -- named entity recognition -- to a larger wildlife trafficking dataset, (i.e., EAGLE briefs spanning multiple countries and species of animals). By applying our methodology to the EAGLE briefs we were able to develop a strategic analysis of wildlife trafficking across West Africa, providing a visualized web service.

\section{Approach}
As shown in Figure~\ref{fig:framework}, our three-phased framework consists of an (1) NLP pipeline, (2) database creation and (3) visualization. A visualization of report data enables investigators to (i) easily identify where these wildlife trafficking incidents are being reported, (ii) what is being trafficked and (iii) by whom. The first phase creates our custom NLP pipeline using spaCy \cite{spaCy2021}, an open-source software library, to read through EAGLE articles and apply named entity recognition (NER) and relation extraction to identify information valuable to wildlife trafficking analysts. The second phase creates a database to organize and store the extracted information. The third phase effectively visualizes the contents of the database with Tableau and provides analysis of the data \cite{tableau}.

\subsection{Data}
Fifty-nine monthly EAGLE briefs beginning January 2017 through December 2021 were downloaded as pdf files. This range of briefs was used as data because they were all written following the same format consisting of a front cover page followed by a one-page summary. This format enabled us to simplify the process of creating our pipeline and extracting data. 
Each EAGLE brief contains both text and images; however we focused on the text-based summary page of each brief as it is dense with high interest information.

\subsection{NLP Pipeline}

\begin{figure*}[ht]
\centering
\subfloat[Base spaCy NER\label{fig:existingNER}]{\includegraphics[width=0.49\linewidth,height=2.2in]{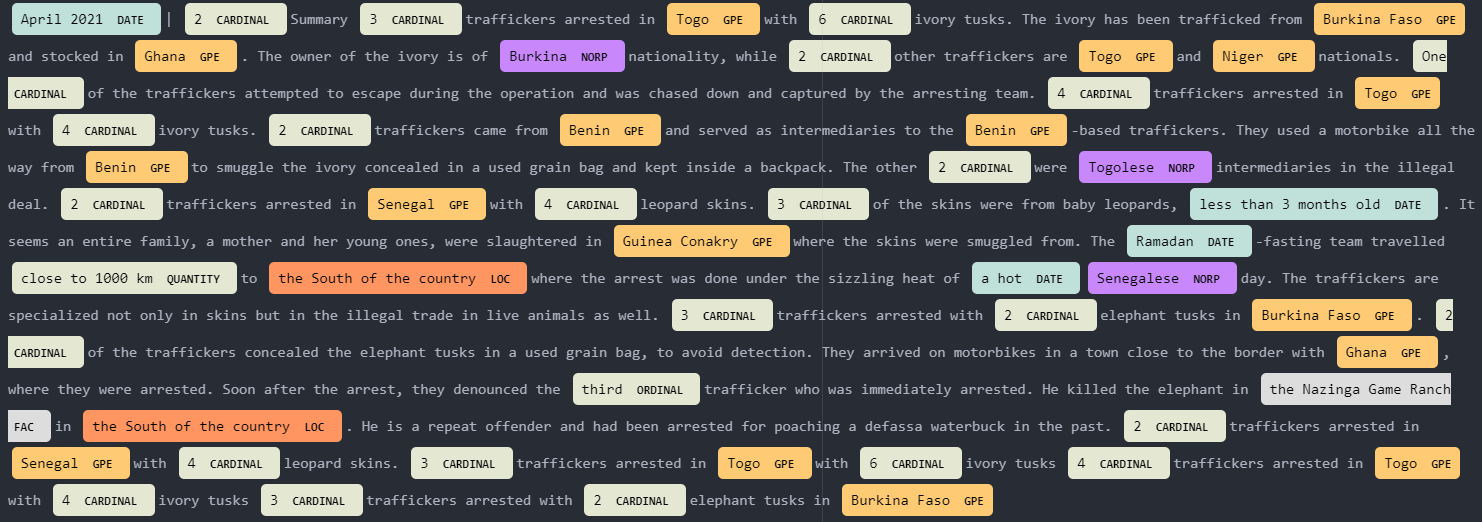}}
\vspace{-5pt}
\subfloat[Our Custom spaCy NER\label{fig:newNER}]{\includegraphics[width=0.49\linewidth,height=2.2in]{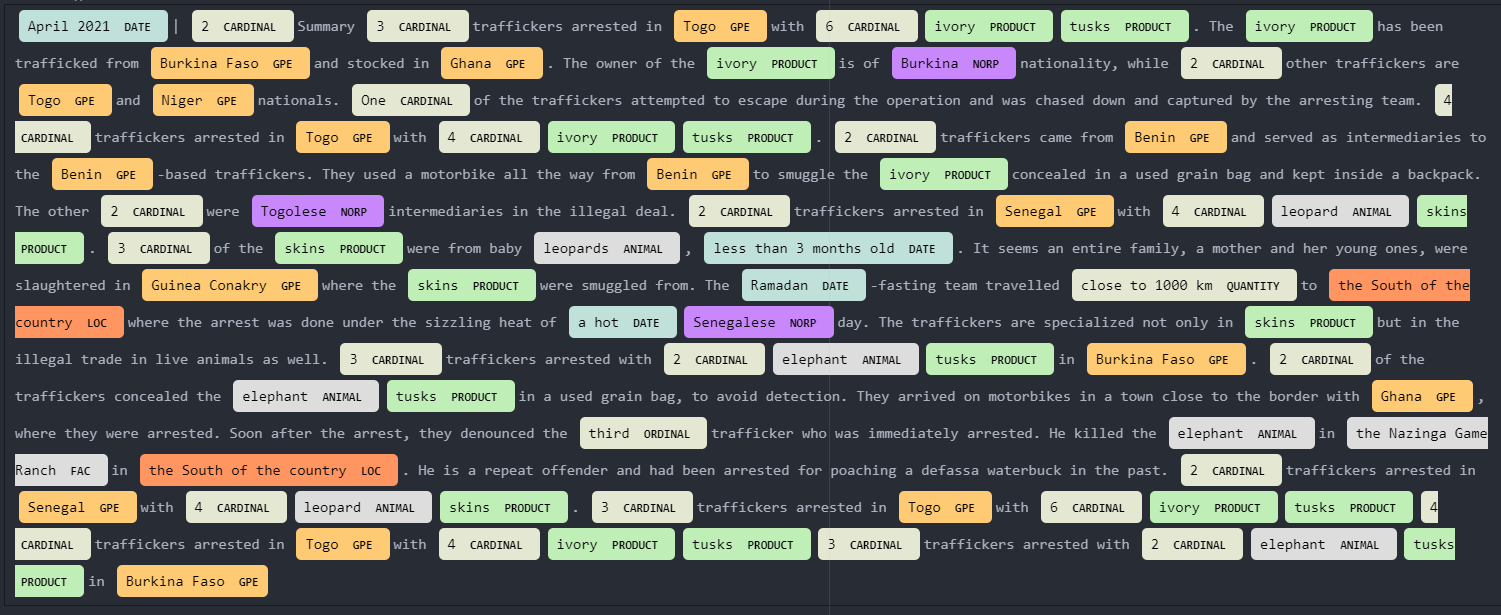}}
\caption{The right figure shows our custom spaCy named entity recognition (NER) results, identifying ANIMAL entities highlighted in gray and PRODUCT entities highlighted in green.}
\label{fig:NER}
\vspace{-10pt}
\end{figure*}

We selected Python’s spaCy to create the NLP pipeline as it has shown to have the best overall performance \cite{shelar2020named}, and contains many pre-trained tools which could be easily customized to fit our project. We used the pre-trained English transformer pipeline (roberta-base) ``en\_core\_web\_trf'' which consisted of a transformer, tagger, parser, attribute ruler, lemmatizer, and NER components. Pipeline output is the tokenized text -– a doc, from which we can access useful information including named entity types, part of speech tags, dependencies, and much more. In its base/original form, this pipeline does not identify certain types of entities of interest namely animals (i.e., fauna). To be able to name custom entities, we added spaCy’s entity ruler to the end of our pipeline (Figure~\ref{fig:framework}).

The entity ruler component allowed us to use a rules-based approach to create our own set of entities to identify in the text: ANIMAL and PRODUCT. Because these entities are finite sets, for each entity type we created a list of those entities. Figure~\ref{fig:existingNER} is a visualization of spaCy’s NER model without our custom named entities on a section of the EAGLE monthly report for April 2021. Types of animals or animal products are not recognized. In contrast, Figure~\ref{fig:newNER} contains the same section of text with the custom named entities. Animal products are highlighted in green while animal species are highlighted in gray.

Starting from a collection of reference lists of animals by common English name from multiple websites, we combined the lists, removed duplicate mentions of the same animal, and reduced them down to 286 animals \cite{listofAnimalNames,allanimals,allanimals1000}. The decision to exclude an animal from our list involved whether the name was too specific. For example, we include the term “sea turtle” but do not include more specific types of sea turtle (e.g.,  loggerhead, kemp's ridley). Some exceptions to this rule were made for animals known to be commonly trafficked, for example the African vs Asian elephant. A handful of rare animals missing from the list were added because they are known to be trafficking targets. We created a list of nine common animal products as well. 
Entities extracted from the EAGLE articles could be in singular or plural form (e.g., tusk and tusks). We utilized the inflect Python package to reformat plural instances \cite{inflect} to achieve label consistency when we insert them into the database.

\subsection{Database}
To design and create the database we worked with wildlife investigators to understand the trafficking trade which resulted in an Entity Relationship Diagram (ERD) for the database schema. To construct the database in our codebase we used the python library SQLite3. Once the database was built, we wrote a function to automatically add the output of the NLP pipeline to the database. The database storing pipeline results is stored to a local comma separated file (.csv), which can be downloaded and opened in MS Excel. This is a desirable format for many domain experts because it is easy to use and view, and can be downloaded from \href{https://wildlifemqp.github.io/Visualizations/}{https://wildlifemqp.github.io/Visualizations/} in the About tab.

\begin{figure}[ht]
\centering
    \includegraphics[width=\linewidth]{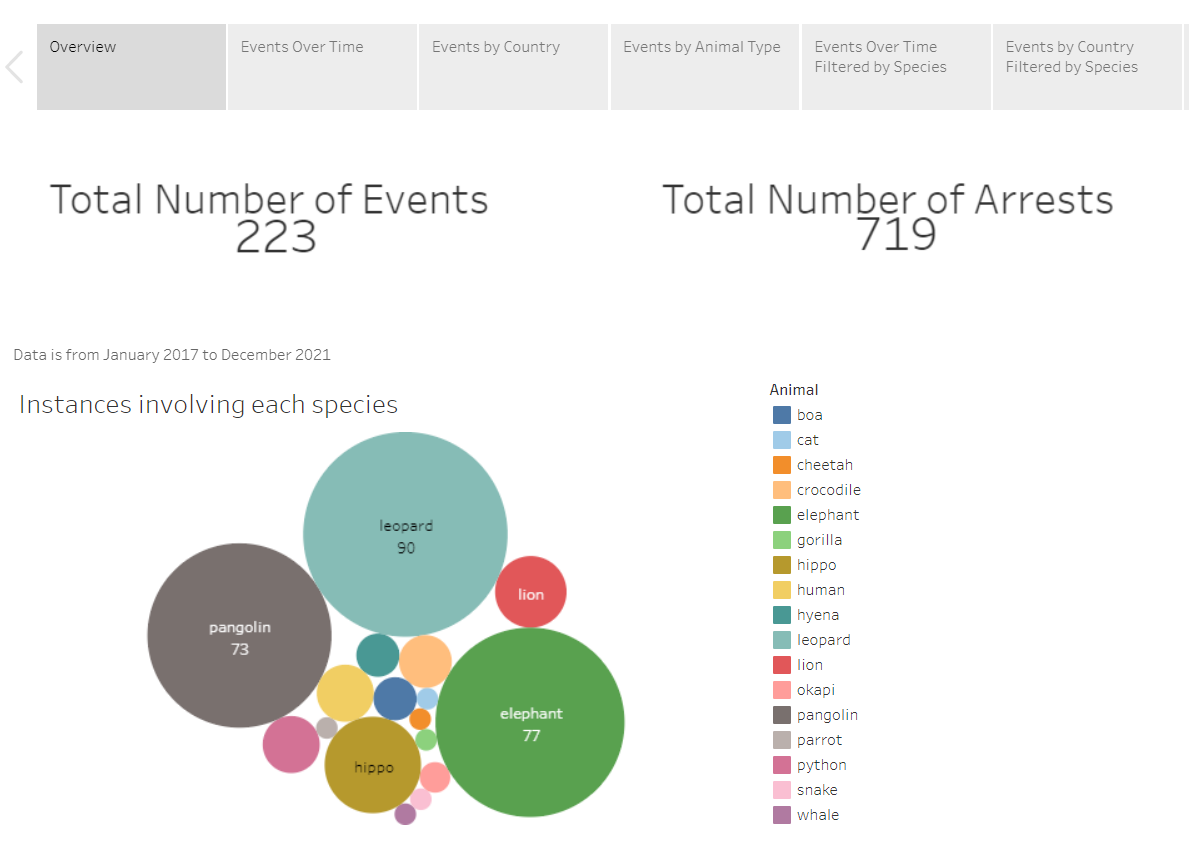}
    \caption{The front page of our website, showing the number of events, arrests and species with the other 9 visualization tabs.}
    \label{fig:website}
\end{figure}

\subsection{Visualization / Website}
Working with subject matter experts on wildlife crime, we selected 10 visualizations to display the data from the database. We choose to use Tableau due to its established reputation, comprehensive features, and visually appealing graphics. Using Tableau Desktop we formatted our visualizations in a story and published our work to Tableau Public. Uploading our visualizations to Tableau Public allowed us to embed the graphics into a website hosted on GitHub. Figure~\ref{fig:website} displays the opening story so viewers could obtain general idea of the data collected and prefaces the other 9 visualizations available. The bubble chart in the figure contains the 5-year data from January 2017 to December 2021, displaying a total of 223 events and 719 total arrests indicating leopards, elephants, and pangolins as the species most seized while being trafficked. The other 9 visualizations can be accessible via the website.

\subsection{Accuracy Analysis of Wildlife Trafficking Event Extraction}
To analyze the accuracy of our NLP pipeline, we compared its results to (i) what a human reader would extract from the same brief as the ground truth, and (ii) outcome extracted by the original spaCy NER as a baseline, which does not contain our event ruler.

To obtain the ground truth, we sampled fifteen briefs in total, three randomly selected from each year between 2017 through 2021. A reader reviewed a brief and manually recorded the information in the same format the pipeline would add an entry to the database. Then the brief was run through our NLP pipeline, and the entries created by the pipeline were compared to the reader’s entries. The fields we compared for each entry were arrest count, country, product name, animal species, quantity, and weight. We used four categories to represent each possible outcome of an entry added by the pipeline to the database: fully correct, partially correct, undetected, and unrelated.

Our NLP pipeline identified 15 fully correct events, 36 partially correct events, and 39 unrelated events. It did not identify 38 out of 85 true events. 
To measure how effective our proposed NLP pipeline is compared with an existing approach, we also ran the original spaCy NER. It identified 0 fully correct events, and remaining retrieved events either only contained country names (i.e., hard to compare them against true events because of limited information) or were unrelated. Overall, our NLP pipeline achieved much better results than existing one (i.e., 15 vs. 0 fully correct event retrieval).

\section{CONCLUDING REMARKS}
Law enforcement authorities and conservation organizations working to combat wildlife trafficking lack a single centralized network for finding and sharing recent data and news. Without an accessible, centralized information source the community of researchers, analysts, and law enforcement are likely to continue finding it extremely tedious and time consuming to identify wildlife trafficking trends and draw inferences. The absence of a single exhaustive wildlife trafficking database prevents experts’ ability to make connections between events and gain a big-picture understanding. Our NLP processing pipeline with its domain specific named entity recognition, expandable database, and interactive website with various illustrative visualizations, serve as a framework for a first step in creating a comprehensive tool for wildlife trafficking data.

Analytical advancements presented herein are not without limitations. The two biggest challenges were the co-reference and relation extraction problems, which were found to be the main sources of error. One potential way to develop the pipeline is to use Prodigy which allows users to manually annotate examples within the dataset to train an AI model \cite{Prodigy}. Prodigy can be used to select named entity recognition training data and NLP relations. This would help ensure the details are connected to the appropriate animal product. This could also be used to link words together such as an animal name and a product type so the model can classify an animal-product entity. Ideally the framework would reach a determined accuracy threshold and eliminate the need for manual intervention. Obvious next steps include extracting additional data from the text such as means of transportation and product details such as price would be useful to users.  Additional data sources should also be included to increase the dataset size, pull in more frequent data, and offer a more comprehensive representation of the true scale of wildlife trafficking.

Despite such limitations, the dashboard created was deployed to the International Union for Conservation of Nature’s PACO program in West and Central Africa, helping member states to aggregate and explore data across countries and scope the nature of wildlife trafficking on the ground. The multitude of graphs and filters offer diverse authorities such as law enforcement, policy makers, and researchers, to enhance the ability of data to inform decision making about a global, pervasive problem that poses risks to diverse ecosystems and people.

\section*{Acknowledgment}
This work was supported by NSF grant IIS-2039951. Any opinions, findings and conclusions or recommendations expressed in this material are the author(s) and do not necessarily reflect those of the sponsors or the U.S. Government.

{
\footnotesize
\bibliographystyle{ieeetr}
\bibliography{ieee}
}

\end{document}